\documentstyle[epsf,aps]{revtex} 
\begin{document}
\newcommand{\p}{\partial}
\newcommand{\ls}{\left(}
\newcommand{\rs}{\right)}
\newcommand{\beq}{\begin{equation}}
\newcommand{\eeq}{\end{equation}}
\newcommand{\beqa}{\begin{eqnarray}}
\newcommand{\eeqa}{\end{eqnarray}}
\newcommand{\bdm}{\begin{displaymath}}
\newcommand{\edm}{\end{displaymath}}
\draft
\title{Origin of subthreshold $K^+$ production in heavy ion collisions
}
\author{
C. Fuchs, Z. Wang,  L. Sehn, Amand Faessler, V.S. Uma Maheswari 
and D.S. Kosov
}
\address{
Institut f\"ur Theoretische Physik der Universit\"at T\"ubingen,\\
Auf der Morgenstelle 14, D-72076 T\"ubingen, Germany
}
\maketitle  
\begin{abstract}
We investigate the origin of subthreshold $K^+$ production in 
heavy ion collisions at intermediate energies. 
In particular we study the influence 
of the pion induced $K^+$ creation processes. We find that this 
channel shows a strong dependence on the size of the system, 
i.e., the number of participating nucleons as well as on the 
incident energy of the reaction. In an energy region between 
1--2 GeV/nucleon the pion induced processes essentially contribute 
to the total yield and can even become dominant in reactions with 
a large number of participating nucleons. Thus we are able to 
reproduce recent measurements of the KaoS Collaboration for 
1 GeV/nucleon Au on Au reactions adopting a realistic 
momentum dependent nuclear mean field.
\end{abstract}
Since strange particles, in particular the $K^+$ meson, 
are considered to be well 
adapted to study the properties of compressed and excited nuclear matter 
produced in heavy ion collisions \cite{AiKo85} strong efforts have 
been made in recent years concerning the measurement of kaon 
observables in intermediate energy reactions \cite{kaos94,fopi95}. 
These subthreshold $K^+$ mesons are predominantly 
produced in the early phase of the nuclear reaction and survive, 
in contrast to, e.g., pions, nearly undistorted by final state 
interactions. Thus, $K^+$ mesons provide a direct source of information 
about the hot and compressed phase and have furthermore found to 
be sensitive on the nuclear equation of state (EOS) 
\cite{huang93,hartnack94}. Transport models, e.g. BUU or QMD calculations, 
have been successful in the description of both, kaon 
\cite{huang93,hartnack94,giessen94,li95b} and pion 
\cite{bass95,fuchs97} yields and spectra.

As a common feature of most theoretical calculations $K^+$ abundances 
and spectra strongly support 
a soft EOS \cite{huang93,hartnack94,li95b} 
whereas pions are less sensitive to the 
nuclear EOS \cite{bass95,fuchs97}. 
Repulsive nuclear forces in general reduce the compression reached 
in the high density phase of the reaction 
and the number of elastic and inelastic scattering 
processes. Hence the kaon yield is reduced as 
well \cite{hartnack94,li95b}. 
However, the reproduction of nuclear flow observables 
requires more repulsive mean fields which in particular 
should take into account the momentum dependence of the 
nuclear interaction \cite{fopi2,fuchs96}. Thus, 
there appears to be a contradiction in model calculations 
with respect to different observables.

The creation mechanism of $K^+$ mesons can be divided into two 
classes of relevant processes, i.e., baryon induced processes 
\beq
BB\longrightarrow BYK^+ 
\label{prim}
\eeq
where the kaon is created via binary baryon--baryon collisions 
($B$ stands either for a nucleon or a $\Delta$--resonance and $Y$ for 
a $\Lambda$ or a $\Sigma$ hyperon, respectively) and processes 
\beq
\pi B\longrightarrow YK^+ 
\label{sec}
\eeq
induced by pion absorption. 

In most previous theoretical studies on subthreshold $K^+$ production 
only the primary processes have been considered, so, e.g., in Refs. 
\cite{huang93,hartnack94,giessen94,li95b}. 
The elementary cross sections used in standard transport calculations 
are those given by Randrup and Ko \cite{ran80} which we also 
apply in the present work. (Although recent 
proton--proton scattering data \cite{cosy96} 
indicate that these might overpredict the $K^+$ production 
near threshold we do not assume the total 
yields too much to be affected by this uncertainty.) 
As a general result it was found that the creation process is strongly 
dominated by the $\Delta$--channels 
\cite{huang93,hartnack94,giessen94,xiong90} 
since the $\Delta$--resonance serves as an 
energy storage for the creation mechanism. 
Furthermore, it was observed that the kaon yield significantly 
depends on the EOS, i.e., a soft EOS enhances the abundancies 
\cite{AiKo85,hartnack94,giessen94,li95b,xiong90}. 
This statement also holds for the relativistic BUU approach of 
\cite{giessen94,li95b} where the repulsion of the model can be interpreted 
in terms of the model dependent effective nucleon mass $m^*$, 
see also Ref. \cite{fuchs97b}. Most of these calculations required 
a weakly repulsive nuclear mean field in order 
to reproduce the experimental kaon data \cite{huang93,hartnack94,li95b}. 
Recent studies \cite{ko95} further indicate that 
the kaon transverse flow might be sensitive on the medium 
dependence of the kaon dispersion relation 
which naturally can be expressed in scalar and vector mean fields. 
However, the magnitude of these fields and a reduction of the 
in-medium kaon mass as well as a possible shift 
of the respective thresholds are 
still a question of current debate \cite{weise93}. Anyway, 
such a shift produced by a kaon potential will equally 
appear in both, baryon and pion induced channels and 
thus not significantly affect their balance. Hence in the present 
work we do not take into account such a potential.

Due to the dominance of the $\Delta$--channel in baryon induced 
processes the importance of the pionic channel is a 
question of general interest. As we show in the 
following the relevance of pion induced processes strongly depends on the 
incident energy and the number of participant nucleons $A_{\rm part}$ 
which is understandable since the number of available pions 
strongly increases with $A_{\rm part}$ \cite{bass95}. 
In Ref. \cite{xiong90} it is claimed that 
these channels contribute less than 25\% to the total kaon yield. 
However, in Ref. \cite{xiong90} only one particular system, i.e., a 
central Ca+Ca collision at 0.8 GeV/nucleon has been considered. In Ref. 
\cite{Li94b} a similar analysis was performed for the Au on Au 
system. However, as in Ref. \cite{xiong90} the $\pi\Delta$ channel 
was not included there and thus the pionic channel was generally 
underpredicted. Further in \cite{Li94b} the influence of the 
initialization on the kaon yield was investigated. We also tested 
this with respect to the treatment of the boost procedure to 
the center-of-mass frame of the nuclei and observed a general 
uncertainity of about 20\% on the total kaon yield. However, the 
balance of the  baryon and pion induced channels was thereby nearly 
unaffected and thus we apply the treatment of Ref. \cite{hartnack94}.

In the present work we study the origin of the $K^+$ mass dependence, i.e. 
the respective multiplicities within the framework of 
Quantum Molecular Dynamics (QMD). As the nuclear mean 
field a soft Skyrme force (K=200 MeV) 
with momentum dependent interactions (SMD) adjusted to the 
empirical nucleon--nucleus optical potential is used 
\cite{hartnack94,fuchs97}. Thus we apply an interaction which is able 
to reasonably reproduce intermediate energy nucleon flow 
data \cite{fopi2}. 
Pions are produced by the decay of $\Delta$(1232) and $N^*$(1440) 
resonances which are the dominant channels in the 
1 GeV domain \cite{bass95}. We take the isospin dependence 
of the respective creation and absorption channels explicitely 
into account. A detailed description can be found in Ref. \cite{fuchs97}. 
Kaons are treated perturbatively, i.e. they do not influence 
the reaction dynamics, similar as described in Ref. \cite{fang93}. 
This method is justified by the small number of kaons and the 
lack of reabsorption due to strangeness conservation. However, they 
are propagated and can undergo elastic rescattering with the 
surrounding nucleons. 
\begin{figure}
\begin{center}
\leavevmode
\epsfxsize = 8cm
\epsffile[160 330 420 530]{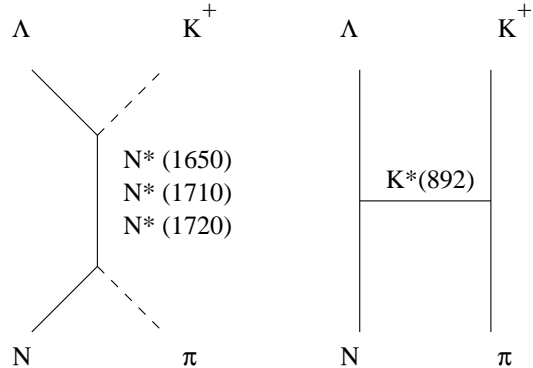}
\end{center}
\caption{Diagrams which contribute to the 
$\pi N \longrightarrow \Lambda K^+$ reaction and which are 
included in the cross section given in Eq. (\protect\ref{siglamb}). 
In the left graph each $N^*$ intermediate state corresponds to 
a seperate diagram.
}
\label{graph}
\end{figure}
In the creation processes, Eqs. (\ref{prim}) and (\ref{sec}), we include 
$\Lambda$ and $\Sigma$ final state hyperons. The elementary cross sections 
for pion induced reactions, Eq. (\ref{sec}), are those given by Tsushima 
et al. \cite{tuebingen1,tuebingen2} 
which are determined in the so-called resonance 
model. In this model the resonances $N^* (1650)(J=\frac{1}{2}^- )$, 
$N^* (1710)(\frac{1}{2}^+ )$, $N^* (1720)(\frac{3}{2}^+ )$ and 
$\Delta (1920)(\frac{3}{2}^+ )$ are included as intermediate states. 
Besides these resonances in the $s$--channel, the $t$--channel 
$K^* (892)$--exchange is included providing a smooth background. 
The relevant coupling constants for baryon--meson vertices 
are thereby determined from the respective decay branching ratios 
of the resonances. This approach is able to reproduce the experimental 
$\pi N \longrightarrow YK^+$ free scattering data. 
The corresponding cross sections including a $\Sigma$ final state 
hyperon, i.e., 
$\pi^+ p \longrightarrow \Sigma^+ K^+$, 
$\pi^- p \longrightarrow \Sigma^- K^+$, 
$\pi^+ n \longrightarrow \Sigma^0 K^+$, 
$\pi^0 n \longrightarrow \Sigma^- K^+$ and 
$\pi^0 p \longrightarrow \Sigma^0 K^+$ 
can be found in Ref. \cite{tuebingen1}. The cross sections 
for the $\Delta$--channel, i.e., 
$\pi^- \Delta^{++} \longrightarrow \Lambda K^+$ and  
$\pi^- \Delta^{++} \longrightarrow \Sigma^0 K^+$, 
$\pi^0 \Delta^{0} \longrightarrow \Sigma^-  K^+$, 
$\pi^+ \Delta^{0} \longrightarrow \Sigma^0  K^+$ and 
$\pi^+ \Delta^{-} \longrightarrow \Sigma^-  K^+$ 
are given in \cite{tuebingen2}. For $\pi^0 p \longrightarrow \Lambda K^+$ 
the resonance model yields \cite{tsushima96}
\beq
\sigma_{\pi^0 p \longrightarrow \Lambda K^+} = 
\frac{1}{2}
\frac{ 0.02279 (\sqrt{s}-1.613)^{0.3894}}{(\sqrt{s}-1.700)^2 + 0.01031} 
\quad {\rm mb}
\label{siglamb}
\eeq
where the diagrams shown in Fig.\ref{graph} have been taken into 
account. In (\ref{siglamb}) $\sqrt{s}$ is given in GeV. From the above 
processes all other isospin channels can be evaluated. 
In contrast to the isospin averaged parametrizations 
of Ref. \cite{cugnon84} these 
cross sections expilcitely distinguish the isospin. 
This fact is of importance with 
respect to the strong isospin 
dependence of pion abundances in systems with small Z/A ratios. 
\begin{figure}
\begin{center}
\leavevmode
\epsfxsize = 10cm
\epsffile[0 100 420 530]{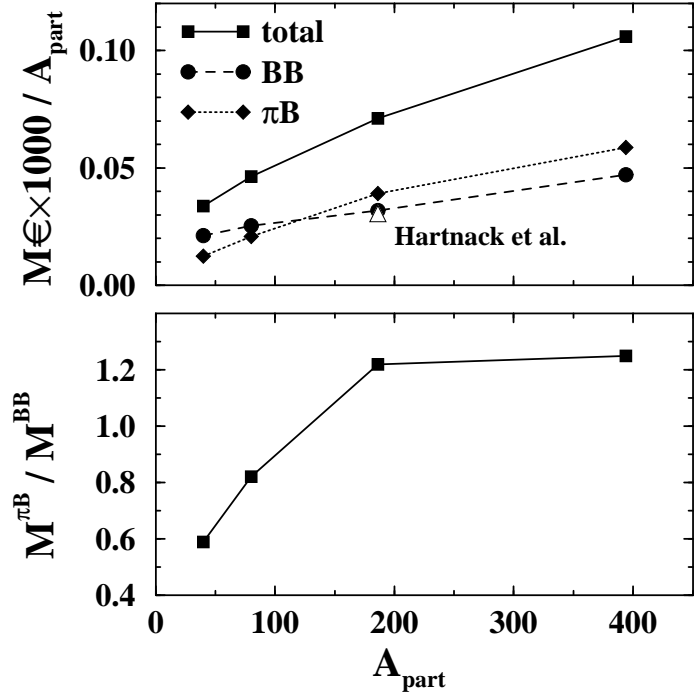}
\end{center}
\caption{$K^+$ mass dependence on the number of 
participating nucleons $A_{\rm part}$. The upper part shows the 
total $K^+$ yield (squares) and the respective contributions 
from baryon--baryon (circles) and pion--baryon (diamonds) 
induced processes in central Ne+Ne, Ca+Ca, Nb+Nb and 
Au+Au collisions at 1 GeV/nucleon. The calculations have been 
performed with a soft momentum dependent Skyrme force. 
In addition the result of Ref. \protect\cite{hartnack94} is shown 
(triangle). The lower part shows the ratio of kaons stemming 
from pion--baryon/baryon--baryon scattering processes.
}
\label{kmass}
\end{figure}
First we examine the mass dependence of the $K^+$ yield. 
In Fig.\ref{kmass} (upper part) 
we show the $K^+$ multiplicities normalized to the number of participating 
nucleons $M_K / A_{\rm part}$ obtained in central collisions (b=0 fm) 
at 1 GeV/nucleon for four 
different symmetric systems, i.e., Ne+Ne, Ca+Ca, Nb+Nb and Au+Au. 
Concerning the baryon induced channel we find a good agreement 
with the corresponding result of Ref. \cite{hartnack94}. We observe a 
general dependence $M_K \sim A^{\tau}_{\rm part}$ ($\tau =1.61$) of the 
total kaon yield on the number of participating nuclei. This 
holds seperately for baryon ($\tau^{BB} =1.40$, 
which is in fair agreement with the results of Ref. \cite{hartnack94}) 
and pion ($\tau^{\pi B} =1.69$) induced processes. 
In the latter case the slope is stiffer and thus we are in total 
closer to experiment ($\tau^{\rm exp} = 1.75\pm0.15$) \cite{kaos2}. 
Consistent with \cite{xiong90} the contribution from $\pi B$ scattering 
is relatively small at low $ A_{\rm part}$. 
The prominent observation 
is, however, that the pion induced channels start to become dominant 
for participant numbers greater than $ A_{\rm part}\sim 150$. This behavior is 
also reflected in the lower part of Fig.\ref{kmass} where the ratio 
$R= M^{\pi B}_K / M^{BB}_K$ of kaons stemming from baryon/pion \
induced processes is shown. 
\begin{table}
\begin{center}
\begin{tabular}{ccccccccc}
      & \multicolumn{2}{c}{Soft} & \multicolumn{2}{c}{Hard}
      & \multicolumn{2}{c}{SMD}  & \multicolumn{2}{c}{HMD}\\ \\
      & M$^{\mathrm{tot}}_{\mathrm{K}}$ 
      & M$^{\mathrm{BB}}_{\mathrm{K}}$ 
      & M$^{\mathrm{tot}}_{\mathrm{K}}$ 
      & M$^{\mathrm{BB}}_{\mathrm{K}}$ 
      & M$^{\mathrm{tot}}_{\mathrm{K}}$ 
      & M$^{\mathrm{BB}}_{\mathrm{K}}$ 
      & M$^{\mathrm{tot}}_{\mathrm{K}}$ 
      & M$^{\mathrm{BB}}_{\mathrm{K}}$  \\
\tableline
Ne+Ne & 3.0  & 2.0  & 2.8  & 1.9  & 1.3  & 0.85 & 1.4  & 0.96  \\
Ca+Ca & 8.9  & 5.2  & 8.5  & 4.9  & 3.7  & 2.0  & 4.3  & 2.1   \\
Nb+Nb & 30.0 & 16.2 & 23.0 & 13.7 & 13.3 & 5.9  & 12.6 & 6.0   \\
Au+Au & 91.0 & 43.2 & 61.5 & 31.0 & 41.8 & 18.6 & 33.5 & 14.6  \\
\end{tabular}
\end{center}
\caption{\label{eos}
Dependence of the $K^+$ production on the nuclear equation of state. 
The same reactions as in Fig. 2 are considered. Skyrme forces 
corresponding to a soft/hard EOS without (Soft/Hard) 
and including momentum dependent interactions (SMD/HMD) are applied. 
M$^{\mathrm{tot}}_{\mathrm{K}}$ is the total multiplicity 
($\times 1000$) and M$^{\mathrm{BB}}_{\mathrm{K}}$ is the 
contribution from baryon induced processes ($\times 1000$). 
}
\end{table}
The influence of the nuclear equation of state on the kaon yield 
is demonstrated in Tab. 1. There we consider the same reactions 
as in Fig.\ref{kmass}, however, applying different EOS's, i.e., a 
soft/hard EOS without and including momentum dependent interactions. 
It is seen that the enhancement of the kaon yield by the pion 
induced channels is a general feature which is rather 
indepent on the particular choice of the nuclear forces. The ratio 
$M^{\pi B}_K / M^{BB}_K$ is more or less the same in all cases. 
As already observed in previous works 
\cite{huang93,hartnack94,li95b,xiong90} the 
kaon yield is generally enhanced using a soft EOS. This effect is 
most pronounced in heavy systems. We further find 
an overall good agreement with the results of Ref. \cite{hartnack94} 
(Fig.9 given therein) concerning the contributions from baryon induced 
$K^+$ production. However, the total kaon yield is 
stronly reduced by the momentum 
dependent interactions (SMD/HMD). Then the stiffness of the EOS 
(soft or hard) is even of minor importance which is due to the fact 
that the repulsion of the interaction originates to most extent 
from their momentum dependence. In the following we restrict the discussion to 
realistic nuclear interactions, i.e., to momentum dependent forces 
(SMD). 
\begin{figure}
\begin{center}
\leavevmode
\epsfxsize = 10cm
\epsffile[0 100 420 530]{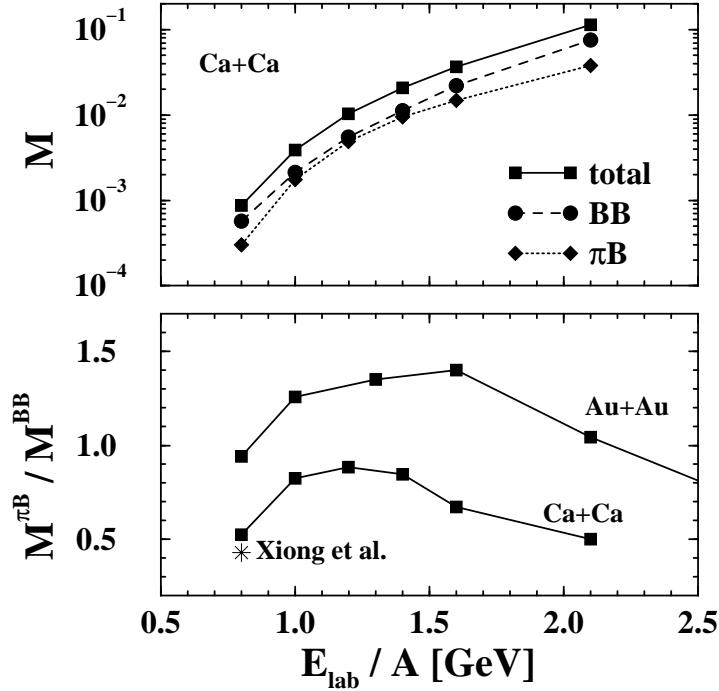}
\end{center}
\caption{Energy dependence of the $K^+$ yield. As in Fig. 2 
the contributions from baryon--baryon (circles) and pion--baryon (diamonds) 
induced processes are compared to the total yield (squares) for 
central (b=0) Ca+Ca 
collisions. The lower part shows the respective ratios obtained 
in central Ca+Ca (b=0) and Au+Au (b=3 fm) collisions. In 
addition the corresponding result of Ref. \protect\cite{xiong90} 
for Ca+Ca at 0.8 GeV/nucleon is shown.
}
\label{kenerg}
\end{figure}
In Fig.\ref{kenerg} the energy dependence of the kaon 
production is considered. 
The upper part shows the kaon multiplicities for a central (b=0 fm) 
Ca+Ca collision at various incident energies and the 
lower part again shows the coresponding ratio $R$. 
It is clearly seen that $R$ exhibits a prominent peak around 
an incident energy of $E_{\rm lab} \sim 1.2$ GeV/nucleon. 
Below 1 GeV/nucleon the pion 
induced processes contribute with 30\% to the total yield which 
coincides with the result of \cite{xiong90} obtained with 
a hard Skyrme force (K=380 MeV). Since we included more repulsive momentum 
dependent iteractions it is reasonable to compare to the hard EOS. 
Due to the lack of the $\Sigma$ channel the total 
yield is underpredicted in \cite{xiong90}, however, the ratios 
are in good agreement. With increasing 
energy the $\pi B$ channel gains more importance. 
At $E_{\rm lab} = 1$ GeV/nucleon its contribution 
lies already at about 45\%. It reaches a maximum of nearly 50\% 
at 1.2 GeV/nucleon and then starts to decrease with energy. A similar 
behavior is observed in a system with large $ A_{\rm part}$, i.e., a central 
(b=3 fm) Au+Au reaction. Here the peak is broadened 
and slightly shifted towards higher energies. 
Above 1 GeV/nucleon the secondary 
processes even start to become dominant. Between an energy range of 
$E_{\rm lab} \sim 1-2$ GeV/nucleon they are responsible for 
55--60\% of the total yield. Furthermore, a 
threshold behavior can be observed 
around 1 GeV/nucleon which can be understood in terms 
of the elementary cross sections. 
Whereas the cross sections for binary baryon--baryon collisions 
$\sigma_{BB\longrightarrow BYK^+} (\sqrt{s})$ are 
more or less linearly increasing functions 
with $\sqrt{s}$ \cite{cugnon84} the respective cross sections 
for the pion induced channels $\sigma_{\pi B\longrightarrow YK^+} (\sqrt{s})$ 
are strongly peaked around $\sqrt{s} \sim 1.8-2$ GeV 
\cite{tuebingen1}. Also a simple 
consideration of the respective thresholds supports these findings. 
E.g. with a final $\Lambda$--hyperon the $\pi\Delta$ 
($\Delta E_{\rm thres} 
= \sum M_{\rm final} - \sum M_{\rm initial} \approx 237$ MeV) 
channel is energetically favoured with respect to the 
$N \Delta$ ($\Delta E_{\rm thres}  \approx 377$ MeV) channel. 
The same holds for the 
$\pi N$ ($\Delta E_{\rm thres}  \approx 530$ MeV) with respect 
to the $NN$ ($\Delta E_{\rm thres} \approx 670$ MeV) 
channel. Providing a sufficiently high amount of pions in the system, 
i.e. large $A_{\rm part}$, the strong contribution 
from $\pi B$ near threshold and its dropping 
down at higher energies is understandable.
\begin{figure}
\begin{center}
\leavevmode
\epsfxsize = 10cm
\epsffile[0 50 400 350 ]{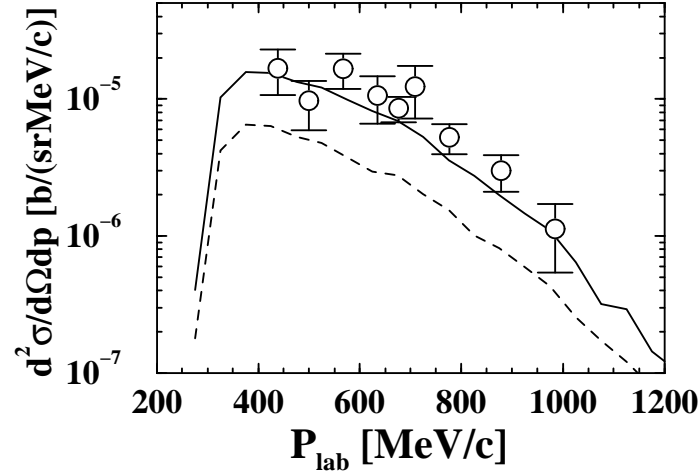}
\end{center}
\caption{$K^+$ spectrum in a Au on Au reaction at 1 GeV/nucleon 
under $40^\circ < \Theta_{\rm lab} < 48^\circ$. The solid line 
indicates the total spectrum and the dashed line the contribution 
only from baryon--baryon induced $K^+$ production. The data are 
taken from Ref. \protect\cite{kaos94}. 
}
\label{kspec}
\end{figure}
\begin{figure}
\begin{center}
\leavevmode
\epsfxsize = 10cm
\epsffile[0 50 400 350 ]{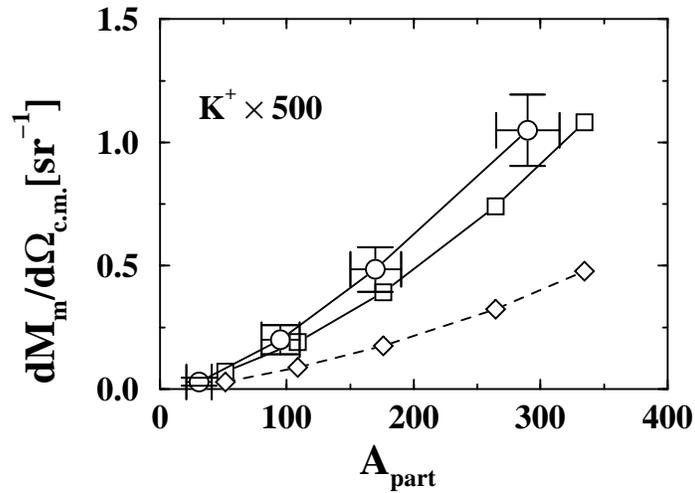}
\end{center}
\caption{$K^+$ multiplicity per center-of-mass solid angle 
for the same reaction as in Fig. 4 as a function of the number 
of participating nucleons. The total (squares) multiplicities 
and the respective contributions from baryon--baryon induced $K^+$ 
production (diamonds) are shown. The data are 
taken from Ref. \protect\cite{kaos94}.
}
\label{kmult}
\end{figure}
\newpage
In Fig.\ref{kspec} the inclusive $K^+$ kinetic spectrum 
measured under $40^\circ < \Theta_{\rm lab} < 48^\circ$ degrees in Au+Au 
reactions at 1 GeV/nucleon is shown as a function of 
laboratory momentum $p_{\rm lab}$. The experimental data are 
those from the KaoS Collaboration \cite{kaos94}. 
The calculations are performed under minimum bias 
condition with an acceptance cut of $0.3< y/y_{\rm proj} <0.6$ 
\cite{kaos94}. We are able to reasonably reproduce the data. 
Consistent with our previous findings, 
Figs.\ref{kmass} and \ref{kenerg}, 
the inclusion only of the $BB$ channel underpredicts 
the data by about a factor of two. The same holds for Fig.\ref{kmult} where 
the $K^+$ multplicities as function of $A_{\rm part}$ 
are compared to the data \cite{kaos94}. 
Here it becomes even more evident that in the heavy system the 
majority of kaons stems from pion induced processes. 
For the total multplicities we find a slight underprediction 
of the data but the general trend is well reproduced, 
in particular the slope of the 
$A_{\rm part}$ dependence at high values of $A_{\rm part}$. 
However, the $K^+$ abundancies found in very recent measurements 
of KaoS with improved statistics \cite{kaos2} seem as well to 
be slightly reduced (by about 20\%) compared to those given 
in Ref. \cite{kaos94}.

To summarize, we have investigated the influence of pion induced 
$K^+$ production in heavy ion collisions. Thereby we applied 
the elementary cross sections determined in the resonance model 
and take both, $\Lambda$ and $\Sigma$ hyperon final states into 
account. With a realistic nuclear mean field, i.e., a soft 
momentum dependent Skyrme force we obtain a good 
agreement with present $K^+$ data from KaoS for both, spectra and 
multiplicities in 1 GeV/nucleon Au on Au 
reactions. Previous calculations which did not take into 
account pion induced processes required an unrealistically 
weak repulsion of the nuclear forces in order to reproduce the 
$K^+$ data, a contradiction which is now resolved. 
We found that the pion induced processes essentially contribute 
to the total yield. This appears to be a general feature which is 
rather independent on the particular choice of the nuclear 
interaction. In heavy systems, i.e., with participant 
numbers greater than about 150 these channels even start to 
become dominant. Furthermore, the pionic channels exhibit a sort 
of a threshold behavior around 1 GeV/nucleon incident energy and 
are most prominent between 1--2 GeV/nucleon which is in agreement 
with the energy dependence of the elementary cross sections 
and the respective thresholds. These 
results further indicate that the measurement of, e.g., 
$K^+ / \pi^+$ correlations is most promising in an energy range 
between 1--2 GeV/nucleon with maximal number of participating 
nucleons.

\end{document}